\documentclass[aps,pre,longbibliography,notitlepage,floatfix,11pt,tightenlines,nofootinbib]{revtex4-1}

\usepackage{enumitem}

\usepackage{bm,amsmath,amssymb,graphicx,stmaryrd,subcaption,tensor}
\graphicspath{ {Figures/} }
\usepackage[abs]{overpic}
\newcommand{\uvc}[1]{\bm{\mathrm{\hat #1}}} 

\newcommand{\Gr}{\varepsilon_{\text{\tiny{G}}}}
\newcommand{\Ar}{\varepsilon_{\text{\tiny{A}}}}
\newcommand{\Br}{\varepsilon_{\text{\tiny{B}}}}
\newcommand{\Sr}{\varepsilon_{\text{\tiny{S}}}}

\newcommand{\bG}{{\bf G}}
\newcommand{\bg}{{\bf g}}
\newcommand{\bR}{{\bf R}}
\newcommand{\br}{{\bf r}}
\newcommand{\bX}{{\bf X}}
\newcommand{\bx}{{\bf x}}
\newcommand{\bA}{{\bf A}}
\newcommand{\ba}{{\bf a}}
\newcommand{\bF}{{\bf F}}
\newcommand{\bU}{{\bf U}}
\newcommand{\bV}{{\bf V}}
\newcommand{\bC}{{\bf C}}
\newcommand{\bB}{{\bf B}}
\newcommand{\bI}{{\bf I}}
\newcommand{\bQ}{{\bf Q}}

\usepackage[format=plain,labelfont={bf,small},textfont=small,justification=raggedright,singlelinecheck=false]{caption}

\begin{document}

\title{Contrasting bending energies from bulk elastic theories}

\author{H. G. Wood}
\email{wood2316@vt.edu}
\affiliation{Department of Biomedical Engineering and Mechanics, Virginia Polytechnic Institute and State University, Blacksburg, VA 24061, U.S.A.}
\author{J. A. Hanna}
\email{hannaj@vt.edu}
\affiliation{Department of Biomedical Engineering and Mechanics, Department of Physics, Center for Soft Matter and Biological Physics, Virginia Polytechnic Institute and State University, Blacksburg, VA 24061, U.S.A.}

\date{\today}

\begin{abstract}
	The choice of elastic energies for thin plates and shells is an unsettled issue with consequences for much recent modeling of soft matter.
	Through consideration of simple deformations of a thin body in the plane, we demonstrate that four bulk isotropic quadratic elastic theories have fundamentally different predictions with regard to bending behavior.  At finite thickness, these qualitative effects persist near the limit of mid-surface isometry, and not all theories predict an isometric ground state.
	We discuss how certain kinematic measures that arose in early studies of rod mechanics lead to coherent definitions of stretching and bending, and promote the adoption of these quantities in the development of a covariant theory based on stretches rather than metrics.
\end{abstract}

\maketitle

\vspace{-0.2in}
\section{Introduction}\label{introduction}

Much recent work in soft matter mechanics has explored the interplay of stretching and bending elasticity in incompatible plates and shells composed of isotropic gels or elastomers, or nematic solids \cite{Efrati09jmps, Na16, GengSelinger12, NguyenSelinger17, Pezzulla15}.
This subject has inspired, among other things, 
 fundamental questions about embeddings \cite{MarderPapanicolaou06, Santangelo09, GemmerVenkataramani13}, 
design principles for shape-programmable materials \cite{Dias11, Modes11, Plucinsky18, Aharoni18}, and insight into the complex shapes of leaves and torn plastic trash bags \cite{AudolyBoudaoud03, Sharon07}.

A work that has been particularly influential on the modeling efforts of several other groups in this community has been that of Efrati and co-workers \cite{Efrati09jmps}, who introduced a formalism for isotropic incompatible elasticity of plates.
This work chose a particular form of elastic energy, quadratic in invariants of Green strain.
Issues have been raised with this strain energy \cite{OshriDiamant17, Hanna18obs} that, while not detracting from the valuable conceptual framework developed in \cite{Efrati09jmps}, have important consequences for any attempts to apply this theory to experiments or use it as a basis for building extended theories.

This note seeks to formalize, flesh out, and amplify a point raised by one of the present authors in a broader paper on plate elasticity \cite{Hanna18obs}, namely that constitutive choices of elastic energy that lead to higher-order effects in the bulk actually have significant \emph{qualitative} effects on derived bending energies, even near the limit of vanishing strains.
This is related to ambiguities in the choice of fundamental quantities for ``small-strain'' expansions.  Such expansions, which are a standard approach in physics, are atypical in the continuum mechanics field, where it is also known that different simple constitutive choices for finite strain measures can have important consequences at moderate to large strains \cite{Batra98, Batra01}.


In the context of two basic deformations of a two-dimensional, compatible plate or flat beam, we compare four isotropic ``quadratic'' elastic energies and their reduced one-dimensional forms within the Kirchhoff-Love framework.
Of these, only one is quadratic in stretch (the ratio of present length to reference length) and leads to a bending energy quadratic in a simple kinematic bending variable favored in some direct theories of rods.
The other bending energies are not only more complicated but lead to qualitatively different predictions for the behavior of the plate.
This very simple calculation allows us to illustrate our point while sidestepping the development of a more general theory featuring shearability, the geometry of a two-dimensional surface, or other complexities.
Yet our findings from this restricted example should be widely applicable.
Additionally, we hope that the notational format we introduce here for Biot strains and similar tensors may prove useful to the soft matter community in developing new theories.

While much has been written on the contrasting limits of small bending and zero stretching in plates and shells, problems in which both stretching and bending are important are less well understood \cite{Steigmann08}, despite being induced by common loading conditions.
For example, pulling and rotating the ends of a sheet requires simultaneous changes in mid-surface lengths and curvatures, as does inflation of a closed vessel under internal pressure.
Even the crumpling regime, in which most of a sheet is approximately isometrically deformed, features significant stretching and bending condensation in ``defects'' whose fundamental structure is still a mystery \cite{Witten07}.
Experiments, both tabletop and microscopic, in soft matter are often performed on moderately thin structures of the type found in engineering applications, where separation of scales is modest.

All of this is to say that bending energy is important, and so it would be a good idea to work with the right one.
The appropriateness of a given energy could be determined by a connection with experimental observations or, if one seeks to predict material-agnostic behaviors, by the simplicity of an otherwise reasonable choice.
However, convenient choices for expressing bulk elasticity as an expansion based on position derivatives 
differ from those that simplify direct models of low-dimensional bodies \cite{Hanna18obs}.

Consider the following question, based on whatever intuition you may have for common elastic materials.
Take a thin plate of such a material and bend it into a cylindrical ring, and fuse the ends so that the midplane is unstretched.  Will the ring:
\begin{enumerate}[noitemsep]
\item Expand to relieve bending and pay some stretching cost (the prediction of an energy quadratic in Almansi or Swainger strains),
\item Stay the same radius (Biot strain),
\item Contract to a smaller radius to relieve bending and pay some stretching (compressive) cost (Green strain)?
\end{enumerate}
It is likely a surprise to many soft matter researchers that they have been working with option 3 for quite a few years.  Perhaps less intuitively obvious is that option 1 is equally undesirable from the point of view of simple direct theories of beams and plates.
A related question is, how do we wish to define a measure of bending, and by extension a bending energy and the concept of a pure stretching deformation, for a thin structure?
One possible answer is that we'd like a bulk energy density quadratic in some measure of strain, a reduced energy density quadratic in the corresponding derived measure of bending, and a linear relationship between this bending measure and the internal moment.  These simultaneous requirements are not trivial to fulfill, and correspond to option 2.

In this note, we consider energies quartic or quadratic in stretch or inverse stretch, built from the quadratic invariants of four types of strain.  All of these invariants are approximately identical for stretches near unity.
The energies can be informally thought of in the following way,
\begin{itemize}[noitemsep]
\item[] Green: quartic, reference
\item[] Biot: quadratic, reference
\item[] Almansi: inverse quartic, present
\item[] Swainger: inverse quadratic, present 
\end{itemize}
Using convected material coordinates, the \emph{covariant} components of Green strain in the coordinate reference basis and those of Almansi strain in the present basis are simply (one half of) the familiar metric differences \cite{GreenZerna92}, quadratic in stretch.  Invariants of these two tensors are formed by raising indices and contracting with the components of the reference or present metrics, respectively.
The Almansi form leads to natural geometric quantities such as curvatures 
of deformed plates and membranes.
The \emph{mixed} components of Biot strain in the reference basis are linear in stretch, and those of Swainger in the present basis are inverse linear.  Another way to express this is that Biot is the finite elastic version of the naive measure given by change in length over reference length, while Swainger is change in length over present length.
We will see that to the usual quadratic order in bending terms, the two ``present'' energies, Almansi and Swainger, give the same bending energy, qualitatively distinct from those arising from Green or Biot.

Biot and similar measures have arisen explicitly or implicitly in both soft matter physics and continuum mechanical works, either as a natural consequence of bead-spring models of molecular mesostructures \cite{OshriDiamant16, SeungNelson88}, from a desire to obtain simple constitutive relations between moment and kinematic variables \cite{OshriDiamant17, KnocheKierfeld11}, or from a recognition that a reduction process will generate certain direct theories of rods \cite{IwakumaKuranishi84, Chaisomphob86, Magnusson01, IrschikGerstmayr09}.
Such direct theories \cite{Antman68-2, Reissner72, WhitmanDeSilva74} for low-dimensional objects are constructed from a convenient choice of kinematic variables without consideration of whether they inherit their form from a bulk elastic theory.

After defining strains and other quantities in Section \ref{definitions}, we derive reduced energies for a restricted geometry in Section \ref{energies}, and examine these for two simple operations in Section \ref{results}, namely the expansion of a closed circular ring, and the extension of a circular arc at fixed radius.  
We finish with some comments in Section \ref{discussion}.

\section{Geometry}\label{definitions}

\begin{figure}[h]
	\includegraphics[width=\linewidth]{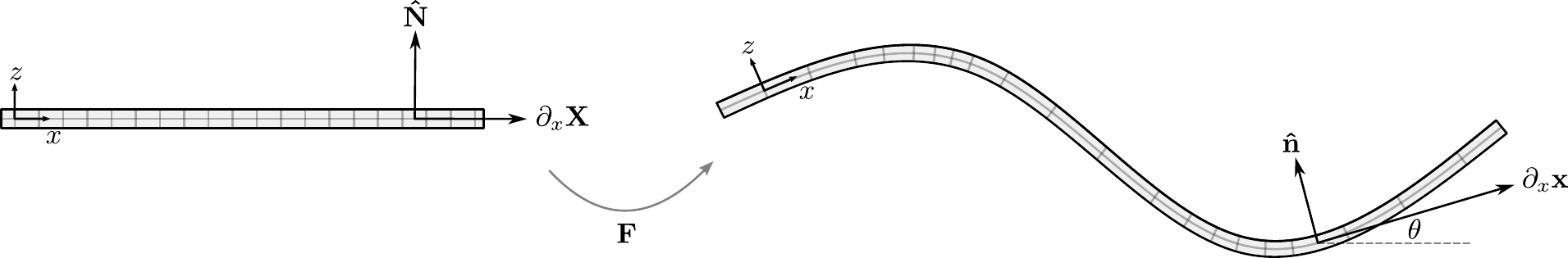}
	\caption{Reference and present configurations of a thin body in the plane.}
	\label{platedef}
\end{figure}

Consider a body in the plane, with uniform thickness $t \ll 1$, parameterized by material coordinates $x$ and $z$ in the axial and thickness directions, respectively.
Its present position is $\br(x,z)$ and that of its mid-line is $\bx(x) = \br(x,0)$.  These two- and one-dimensional objects have natural coordinate bases $\bg_i=\partial_i\br$ and $\ba_x = \partial_x\bx = \bg_x(x, 0)$, respectively.  The mid-line has a unit normal $\uvc{n}$.
The body also has corresponding reference quantities $\bR$, $\bX$, $\bG_i$, $\bA_x$, and $\uvc{N}$.
For simplicity we take $x$ to be the reference arc length, so that $\bA_x$ is a unit vector.
Reciprocal bases of inverse tangents are defined such that $\bg^i\cdot\bg_j = \delta^i_j$ and $\bG^I\cdot\bG_j = \delta^I_j$, with the definitions of $\ba^x$ and $\bA^X$ following in the obvious way.  Capitalization on upper indices is simply a reminder that such indices appearing on reference objects are not raised with the present inverse metric; implied summation ignores case \cite{Hanna18obs}.  Note that $x$ and $z$ or $X$ and $Z$ will always serve as specific material indices and not free or summable indices.
Metric and inverse metric components can be defined using the tangent vectors and their inverses, but we will not need them.
The deformed mid-line is conveniently described using a stretch $\Lambda(x)$ and a tangential angle $\theta(x)$, so that the tangent and normal have Cartesian representations
$\ba_x = \partial_x\bx = \Lambda\begin{pmatrix}\cos\theta \\ \sin\theta \end{pmatrix}$ and $\uvc{n} = \begin{pmatrix}-\sin\theta \\ \cos\theta \end{pmatrix}$. 
Figure \ref{platedef} is a schematic of the reference and present configurations.

The deformation gradient $\bF = \bg_i\bG^I$ maps line elements from reference to present configurations, $\bF \cdot d\bR = d\br$. 
It will be used along with its transpose
$\bF^T = \bG^I\bg_i$ to construct four symmetric tensors.  The right and left Cauchy-Green deformation tensors are $\bC = \bF^T \cdot \bF$ and $\bB = \bF \cdot \bF^T$.  The right and left stretch tensors are defined by the decompositions 
$\bF = \bQ\cdot\bU = \bV\cdot\bQ$, where $\bQ$ is a rotation tensor.  The four strains of interest involve $\bC$, $\bU$, the inverses $\bB^{-1}$ and $\bV^{-1}$, and the identity $\bI$, where the inverse of a symmetric tensor $\bf{T}$ is defined such that $\bf{T}\cdot\bf{T}^{-1} = \bf{T}^{-1}\cdot\bf{T} = \bI$.

For exemplary --- not endorsement --- purposes, we consider the idealized constrained kinematics of the Kirchhoff-Love assumptions 
and associated standard but non-rigorous procedures for deriving a reduced energy.  
The first Kirchhoff-Love assumption tells us that the $x$ and $z$ directions correspond to principal stretches $\Lambda_x(x,z)$ and $\Lambda_z(x,z)$.
We will write things in somewhat non-traditional format using mixed reference or present bases in principal directions.  As with the non-coordinate orthonormal principal direction bases commonly used in continuum mechanics \cite{Kellynotes}, the invariants can be read off easily from these mixed components, but we believe this representation will be more amenable to future theoretical generalizations in convected coordinates.\footnote{We note one disadvantage of this asymmetric representation of symmetric tensors. When used in concert with our adoption above of the traditional transpose notation for the deformation gradient (the latter is unfortunately entrenched in continuum mechanics, and serves to indicate which tensor legs the inner product refers to), one may get the mistaken impression that it might make sense to write transposes of these symmetric tensors.  To be consistent, one should either replace the transpose notation, or symmetrize the bases in expressions such as (\ref{rightcg}-\ref{Swaingerstrain}).  We emphasize that these expressions are not general, but work for our restricted situation.}

Thus, in the restricted situation we are concerned with, the relevant deformation tensors may be written as
\begin{align}
	\text{right Cauchy-Green}&\quad \bC \quad= \Lambda_x^2\bG_x\bG^X + \Lambda_z^2\bG_z\bG^Z, \label{rightcg} \\
	\text{inverse left Cauchy-Green}&\quad \bB^{-1} = (1/\Lambda_x^2)\bg_x\bg^x + (1/\Lambda_z^2)\bg_z\bg^z, \label{ileftcg}\\
	\text{right stretch}&\quad \bU \quad= \Lambda_x\bG_x\bG^X + \Lambda_z\bG_z\bG^Z, \label{rightstretch}\\
	\text{inverse left stretch}&\quad \bV^{-1} = (1/\Lambda_x)\bg_x\bg^x + (1/\Lambda_z)\bg_z\bg^z, \label{ileftstretch}
\end{align}
and the corresponding strain tensors as
\begin{align}
	\text{Green}&\quad  \tfrac{1}{2}(\bC - \bI) \quad = \tfrac{1}{2}\left(\Lambda_x^2 - 1\right)\bG_x\bG^X + \tfrac{1}{2}(\Lambda_z^2 - 1)\bG_z\bG^Z, \label{GLstrain}\\
	\text{Almansi}&\quad  \tfrac{1}{2}(\bI - \bB^{-1}) = \tfrac{1}{2}\left(1 - 1/\Lambda_x^2 \right)\bg_x\bg^x + \tfrac{1}{2}\left(1 - 1/\Lambda_z^2 \right)\bg_z\bg^z, \label{EAstrain}\\
	\text{Biot}&\quad  \quad \bU - \bI  \quad\,\,  =\;\;\;
	\left(\Lambda_x - 1\right)\bG_x\bG^X + (\Lambda_z - 1)\bG_z\bG^Z, \label{Biotstrain}\\
	\text{Swainger}&\quad  \quad \bI - \bV^{-1} \; = \;\;\; \left(1 - 1/\Lambda_x \right)\bg_x\bg^x + \left(1 - 1/\Lambda_z \right)\bg_z\bg^z. \label{Swaingerstrain}
\end{align}
The Green and Almansi tensors will be more familiar to physicists as $\gamma_{ij}\bG^I\bG^J$ and $\gamma_{ij}\bg^i\bg^j$, using the metric differences $2\gamma_{ij}  = \bg_i\cdot\bg_j - \bG_i\cdot\bG_j$ , a form that holds for more general deformations.

A two-dimensional, isotropic, quadratic energy in terms of a nonspecific 
strain tensor $\bm{\epsilon}(\br)$ will take the form
\begin{equation}
	\int \! d\bar{A} \left[c_1\left(\text{Tr}(\bm{\epsilon})\right)^2 + c_2\text{Tr}\left(\bm{\epsilon}^2\right)\right],
	\label{energy}
\end{equation} 
where $d\bar{A} = 1dxdz$ is the simple reference volume measure and the $c_i$ are moduli.
The first Kirchhoff-Love assumption allows for determination of the transverse strain as proportional to the axial strain \cite{Efrati09jmps},  
$\tensor{\epsilon}{^z_z} = f(c_1,c_2)\tensor{\epsilon}{^x_x}$ , 
and thus rewriting of the energy\footnote{This form implies that, for our restricted situation, a quadratic energy built with Biot invariants will be equivalent to an incompressible neo-Hookean energy.}
\begin{equation}
	\int \! dx \, \mathcal{E} = \int \! dx \int_{-t/2}^{t/2} dz \, \tilde{Y}\left(\tensor{\epsilon}{^x_x}\right)^2 ,
	\label{rewrittenenergy}
\end{equation}   
with some suitably defined modulus $\tilde{Y}$.
After this, the second Kirchhoff-Love assumption can be invoked, setting the geometry of the present configuration as $\br(x,z) = \bx(x) + z\uvc{n}(x)$ 
and allowing us to express the axial stretch $\Lambda_x(x,z)$ in terms of the mid-line stretch $\Lambda(x) = \Lambda_x(x,0)$ and angle $\theta(x)$ 
\cite{IrschikGerstmayr09},
\begin{equation}
	\Lambda_x(x,z) = \Lambda(x) - z \partial_x \theta ,
	\label{2dlambda}
\end{equation}
so that we may integrate to explicitly obtain the density $\mathcal{E}$.

\section{Four quadratic energies}\label{energies}

We may now evaluate the forms of the energy density $\mathcal{E}$ from \eqref{rewrittenenergy} that take as input the axial component of the four strains (\ref{GLstrain}-\ref{Swaingerstrain}) containing the bulk axial stretch \eqref{2dlambda}.  For each density, this will result in quadratic terms involving one of the four corresponding mid-line strains $\varepsilon(\bx)$, namely 
\begin{align}
	\text{Green}&\quad \Gr = \tfrac{1}{2}\left(\Lambda^2-1\right), \label{1dGL}\\
	\text{Almansi}&\quad \Ar = \tfrac{1}{2}\left(1-1/\Lambda^2\right), \label{1dEA}\\
	\text{Biot}&\quad \Br = \quad\; \Lambda-1, \label{1dBiot}\\
	\text{Swainger}&\quad \Sr\, = \quad\; 1-1/\Lambda, \label{1dSwainger}
\end{align}
and one of three bending measures,
\begin{enumerate}[noitemsep]
	\item Almansi or Swainger: The present arc length derivative of the tangential angle, which generates the invariant (Frenet) curvature $\partial_s\theta =\partial_x\theta/\Lambda$.  In the Almansi context, this can be thought of as applying the inverse present metric to the covariant component of the curvature tensor, $\left(\ba^x\cdot\ba^x\right)b_{xx}$, where $b_{xx} = \partial_x\ba_x \cdot \uvc{n} = \Lambda\partial_x\theta$.	
	\item Biot: The material derivative of the tangential angle $\partial_x\theta$,
	\item Green: $\Lambda\partial_x\theta$, which can be thought of as obtained by applying the trivial inverse reference metric $\bA^X\cdot\bA^X$ to $b_{xx}$, viewed now as a component of some unnamed tensor.
\end{enumerate}
While the four mid-line strains all linearize to $\Lambda - 1$ when $\Lambda$ is close to unity, the three bending measures have qualitatively different forms.  They are either inversely linearly dependent, independent, or linearly dependent on $\Lambda$.
Interestingly, the two ``present'' energies agree on the definition of bending energy.

We will derive expansions to quadratic order in strain $\varepsilon$ and thickness $z\partial_x\theta \le t\partial_x\theta$, dropping any pure or mixed terms of higher order.
  Each expansion is different because the strains $\varepsilon$ have different dependencies on stretch.  All have the same structure at quadratic order in strain and thickness, namely a stretching term linear in thickness and quadratic in strain, and a bending term cubic in thickness and quadratic in a bending measure.  Although it is a bit redundant, we write out each expansion to emphasize differences at higher order and in terms linear in thickness that, while integrating to zero for a symmetric plate, may be relevant in shells.
The Green density is
\begin{align}
	\mathcal{E}_{\text{\tiny{Green}}} &= \int_{-t/2}^{t/2} dz \, \tilde{Y}\left(\tfrac{1}{2}\left[\left(\Lambda - z \partial_x \theta\right)^2 - 1\right]\right)^2 \nonumber \\
	&= \int_{t/2}^{t/2} dz \, \tilde{Y}\left(\Gr^2 - 2z\Gr\Lambda\partial_x \theta + z^2\left(\Lambda \partial_x \theta\right)^2\right) + O\left(\text{cubic}\right) 
 \nonumber \\
	&\approx t\tilde{Y}\left[\tfrac{1}{2}(\Lambda^2 - 1)\right]^2 + \tfrac{1}{12}t^3\tilde{Y}\left(\Lambda\partial_x\theta\right)^2. 
	\label{Greenenergy}
\end{align}
The Almansi density is
\begin{align}
	\mathcal{E}_{\text{Almansi}} &= \int_{-t/2}^{t/2} dz \, \tilde{Y}\left(\tfrac{1}{2}\left[1 - \frac{1}{\left(\Lambda - z \partial_x \theta\right)^2}\right]\right)^2, \nonumber \\
	&= \int_{-t/2}^{t/2} dz \, \tilde{Y} \left(\Ar^2 - 2z\Ar\partial_x\theta/\Lambda + z^2\left(\partial_x\theta/\Lambda\right)^2\right) + O\left(\text{cubic}\right) \nonumber \\
	&\approx t\tilde{Y}\left[\tfrac{1}{2}\left(1 -1/\Lambda^2\right)\right]^2 + \tfrac{1}{12}t^3\tilde{Y}\left(\partial_x \theta/\Lambda\right)^2. \label{Almansienergy}
\end{align}
The Biot density is special in that, given the prior assumptions in the approach we've taken, it is exact at quadratic order,
\begin{align}
	\mathcal{E}_{\text{Biot}} &= \int_{-t/2}^{t/2} dz \, \tilde{Y}\left(\Lambda - z \partial_x \theta - 1\right)^2 \nonumber \\
	&= \int_{-t/2}^{t/2} dz \, \tilde{Y}\left(\Br^2 - 2z\Br\partial_x \theta + z^2(\partial_x\theta)^2\right) \nonumber \\
	&= t\tilde{Y}\left(\Lambda - 1\right)^2 + \tfrac{1}{12}t^3\tilde{Y}\left(\partial_x \theta\right)^2. \label{Biotenergy}
\end{align}
The Swainger density is 
\begin{align}
	\mathcal{E}_{\text{Swainger}} &= \int_{-t/2}^{t/2} dz \, \tilde{Y}\left(1 - \frac{1}{\Lambda - z \partial_x \theta}\right)^2 \nonumber \\
	&= \int_{-t/2}^{t/2} dz \, \tilde{Y}\left(\Sr^2 - 2z\Sr \partial_x \theta/\Lambda + z^2\left(\partial_x \theta/\Lambda\right)^2\right) + O\left(\text{cubic}\right) \nonumber \\
	&\approx t\tilde{Y}\left(1 - 1/\Lambda\right)^2 + \tfrac{1}{12}t^3\tilde{Y}\left(\partial_x \theta/\Lambda\right)^2. \label{Swaingerenergy}
\end{align}

In this simple context, one can see upon close inspection that the non-Biot bending terms could be rewritten as Biot plus a higher-order correction.
This clumsy additional step would be less apparent in a general treatment that does not admit a simple description in terms of a single stretch and angle.

\section{Two simple deformations}\label{results}

Here we demonstrate the individual and combined behavior of the stretching and bending terms $\mathcal{E}_S$ and $\mathcal{E}_B$ from the four energies (\ref{Greenenergy}-\ref{Swaingerenergy}).
The expansion of a closed circular ring and the extension of a circular arc at fixed radius will be considered.  

\begin{figure}[h]
	\includegraphics[width=0.6\linewidth]{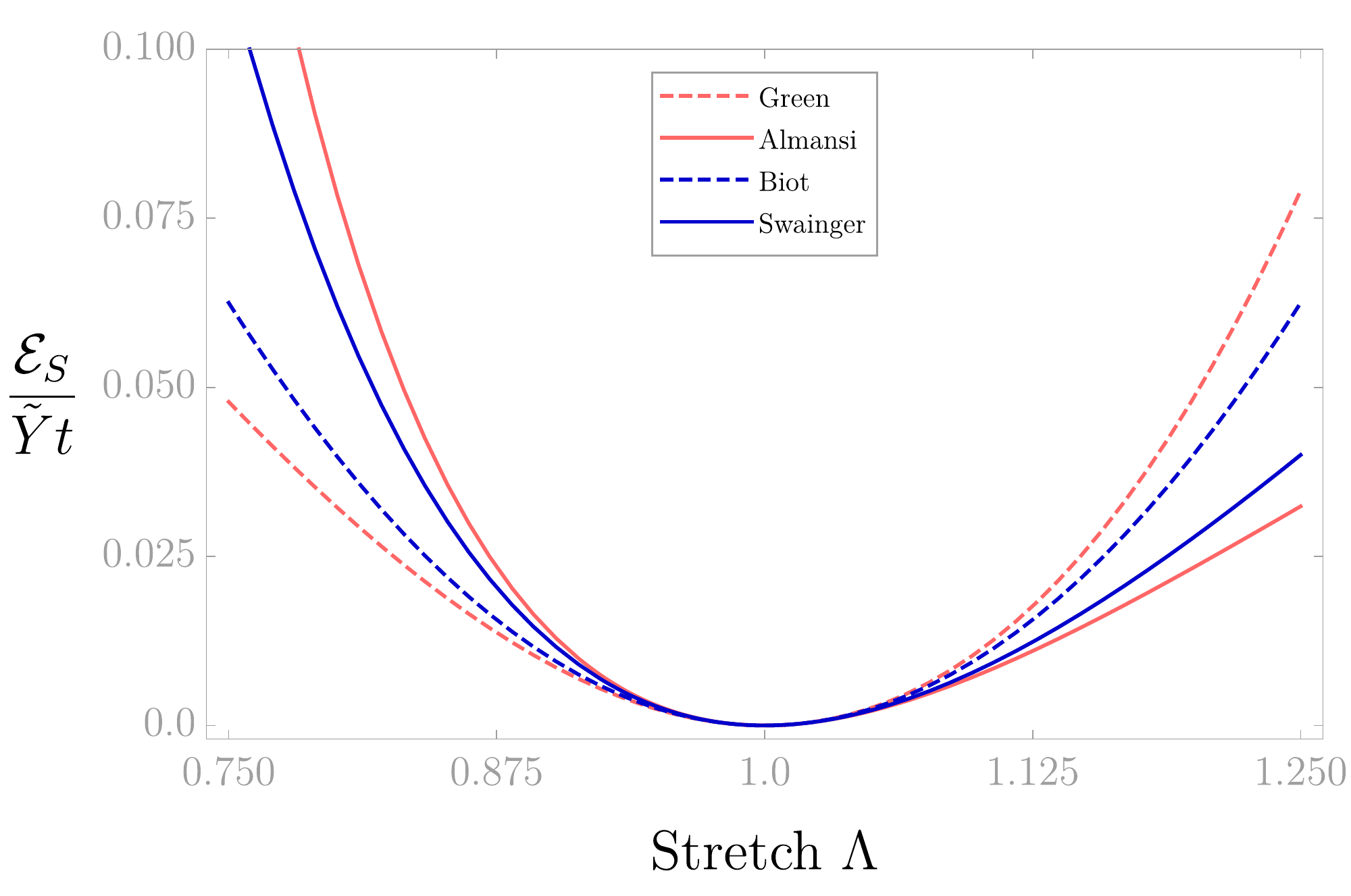}
	\caption{Stretching energy density $\mathcal{E}_S$ as a function of mid-line stretch $\Lambda$. Green and Biot are bounded as $\Lambda\rightarrow0$ while Almansi and Swainger are bounded as $\Lambda\rightarrow\infty$.}
	\label{StretchingEnergyPlot}
\end{figure}

The stretching contents of the four energies for a general deformation are shown in Figure \ref{StretchingEnergyPlot}.
As is well known, the curves share a minimum, and thus are tangent to each other at $\Lambda = 1$, and remain close nearby.
Important differences are found at large deformations, irrelevant to the present discussion of quadratic energies.  The ``reference'' Green and Biot energies are bounded in compression ($\Lambda \rightarrow 0$) while the ``present'' Almansi and Swainger energies are bounded in tension ($\Lambda \rightarrow \infty$), reflecting the fact that the respective measures are normalized using reference or present quantities.

\begin{figure}[h]
	\begin{subfigure}{0.3\linewidth}
		\centering
		\includegraphics[width=1\linewidth]{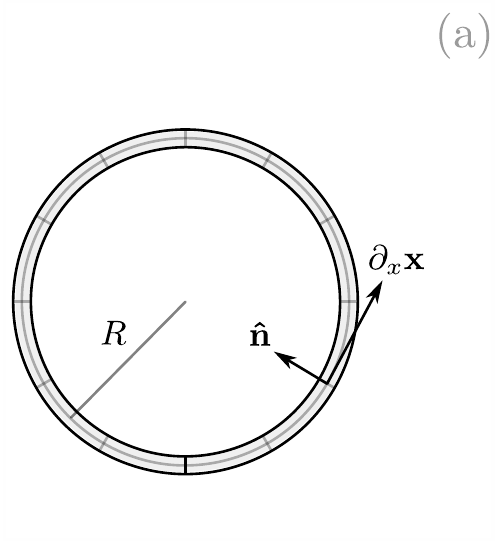}
		\label{expandingring}
	\end{subfigure}%
	\begin{subfigure}{0.7\linewidth}
		\centering
		\includegraphics[width=0.9\linewidth]{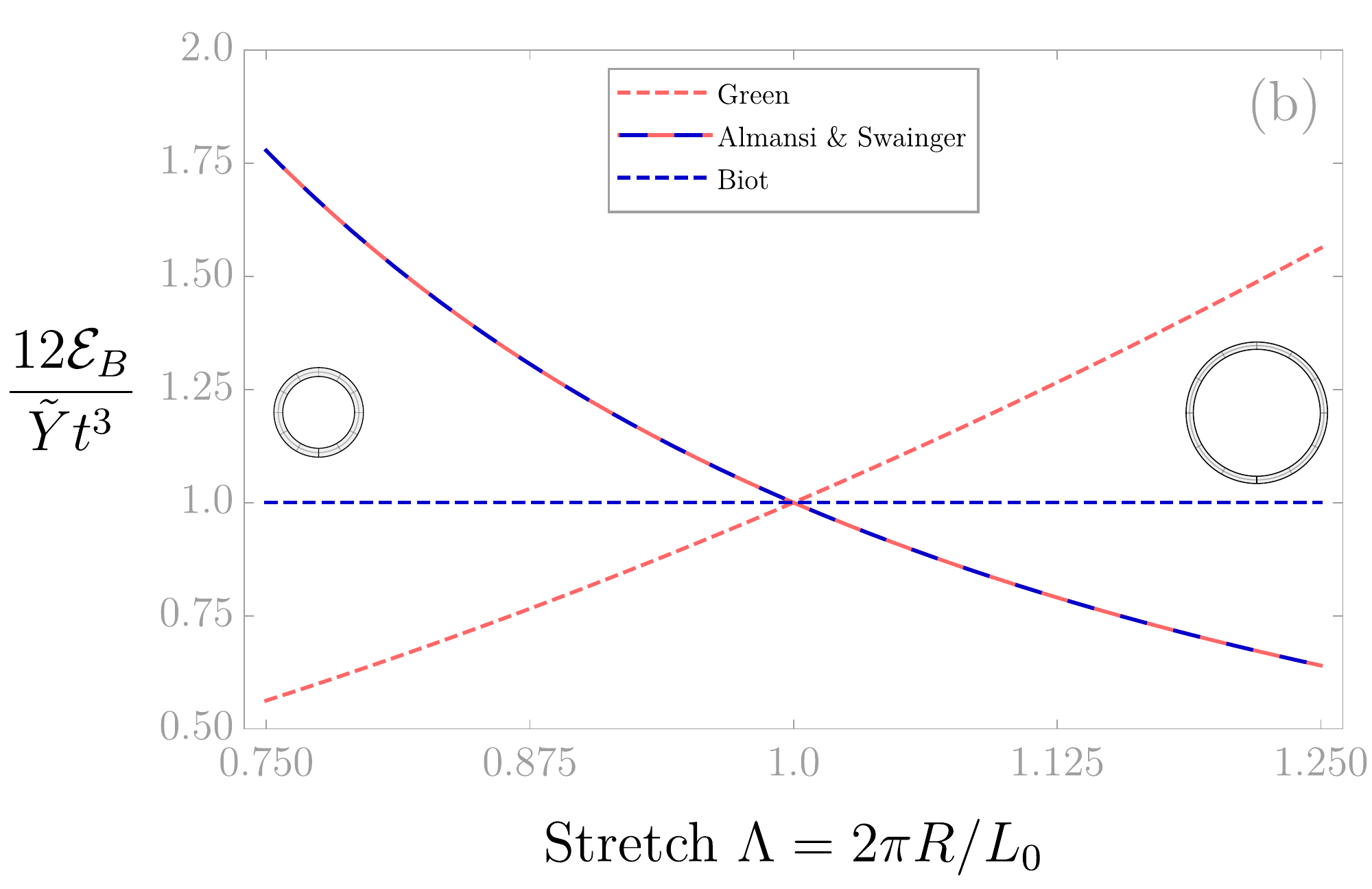}
		\label{expandingenergy}
	\end{subfigure}
	\caption{ {\bf{(a)}} Geometry of a closed ring of naturally flat material. {\bf{(b)}} Bending energy densities as a function of ring radius $R$.  Biot is independent of $R$ and thus stretch $\Lambda$ in this geometry.  Green $\rightarrow 0$ as $R \rightarrow 0$ while Almansi/Swainger $\rightarrow 0$ as $R \rightarrow \infty$.}
	\label{ExpandingRadiusFigure}
\end{figure}

Consider a thin plate or beam, a body whose rest state is flat and of length $L_0$.  The body is bent into a circular cylinder or ring, with its ends fused together (Figure \ref{ExpandingRadiusFigure}).
The resulting bending energies from the four models have qualitatively different dependencies on ring radius $R$.  
The material derivative of the angle, $\partial_x \theta$, and thus the Biot bending energy, are independent of radius.  Expansion or contraction of a ring is ``pure stretching'' in the Biot model, a definition preferred by Antman \cite{Antman05}.  
The identical Almansi and Swainger bending energies are given by the curvature, so decrease with increasing radius, approaching zero as the radius tends to infinity.
The Green bending energy does the opposite --- it approaches zero with the radius, and increases without bound as the ring expands and its curvature vanishes.
These qualitative differences in bending energy appear as zero, negative, or positive slopes at unit stretch, when the mid-line strain vanishes.  The curves are nowhere tangent to each other.

\begin{figure}[h]
	\begin{subfigure}{0.3\linewidth}
		\centering
		\includegraphics[width=1\linewidth]{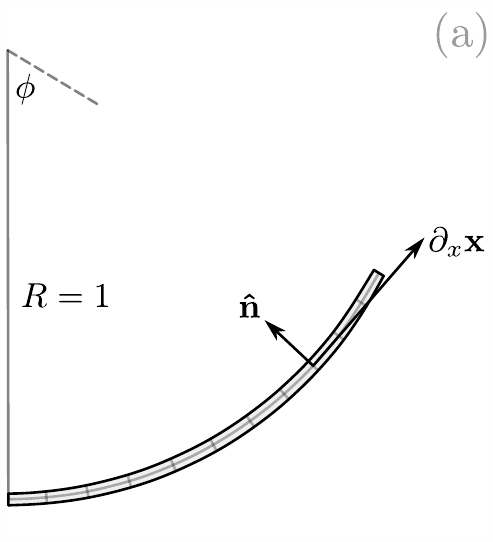}
		\label{arcextension}
	\end{subfigure}%
	\begin{subfigure}{0.7\linewidth}
		\centering
		\includegraphics[width=0.9\linewidth]{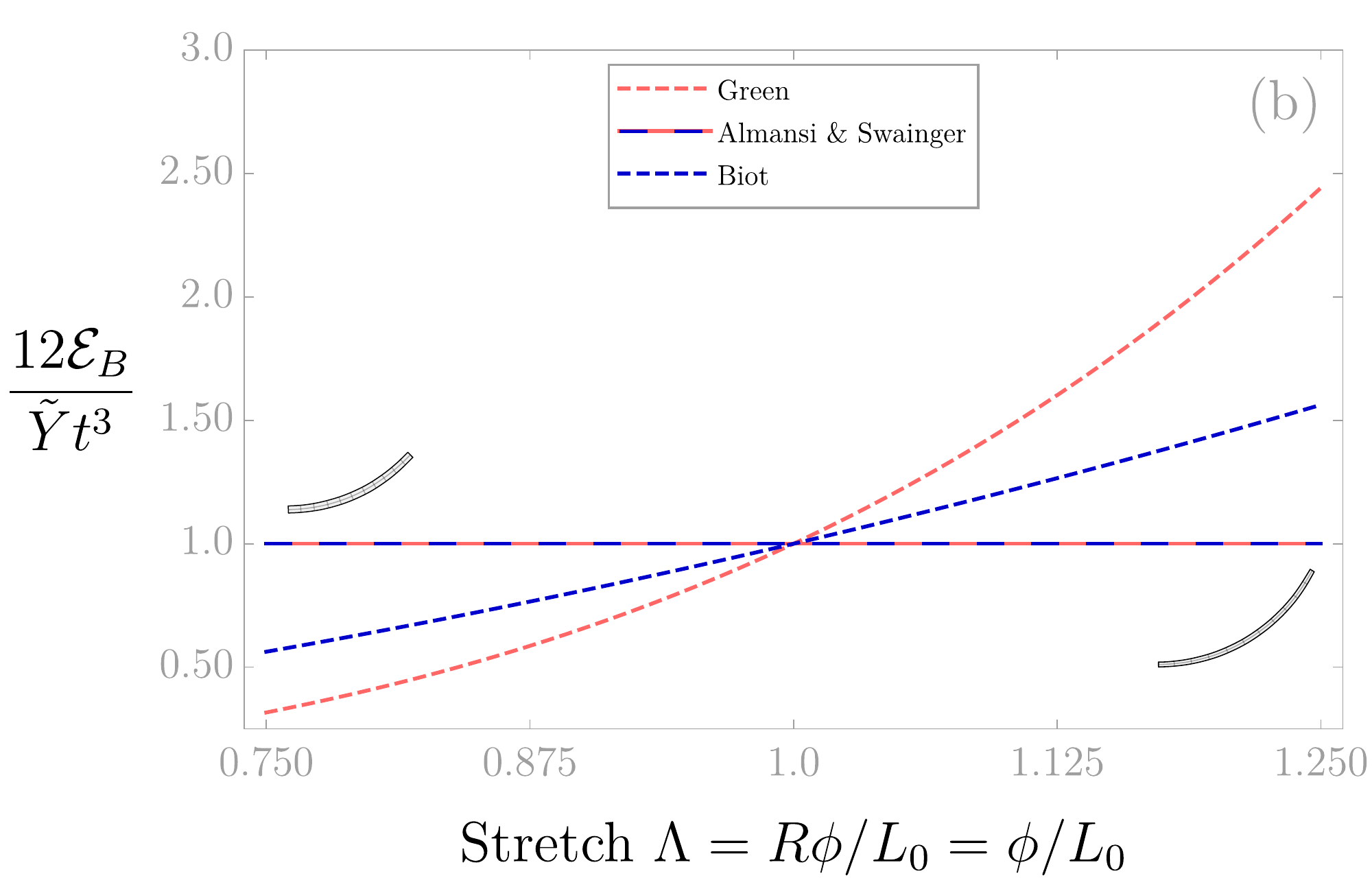}
		\label{extensionenergy}
	\end{subfigure}
	\caption{ {\bf{(a)}} Geometry of an open arc of naturally flat material. {\bf{(b)}} Bending energy densities as a function of subtended angle $\phi$.  Almansi/Swainger is independent of $\phi$ and thus stretch $\Lambda$ in this geometry.  Green and Biot $\rightarrow 0$ as $\phi \rightarrow 0$.}
	\label{ArcExtensionFigure}
\end{figure}

Consider the same plate bent into a circular arc of unit radius, and the dependence of the bending energies on the length of this arc or, equivalently, its subtended angle $\phi$ (Figure \ref{ArcExtensionFigure}).  
The curvature, and thus the identical Almansi and Swainger bending energies, are independent of the subtended angle, making extension or compression of any arc at fixed radius ``pure stretching'' in either of these models.
The Biot and Green bending energies increase without bound with subtended angle, and approach zero as the arc shortens to zero length.

What is ``pure stretching'' in the Green model?  This would be any combination of changes in length and curvature that preserve the value of $\Lambda\partial_x\theta = \Lambda^2\partial_s\theta$.
In terms of circular arcs, this could be achieved by simultaneous radial expansion and axial compression, or radial shrinkage and axial extension.

\begin{figure}[h]
	\includegraphics[width=0.6\linewidth]{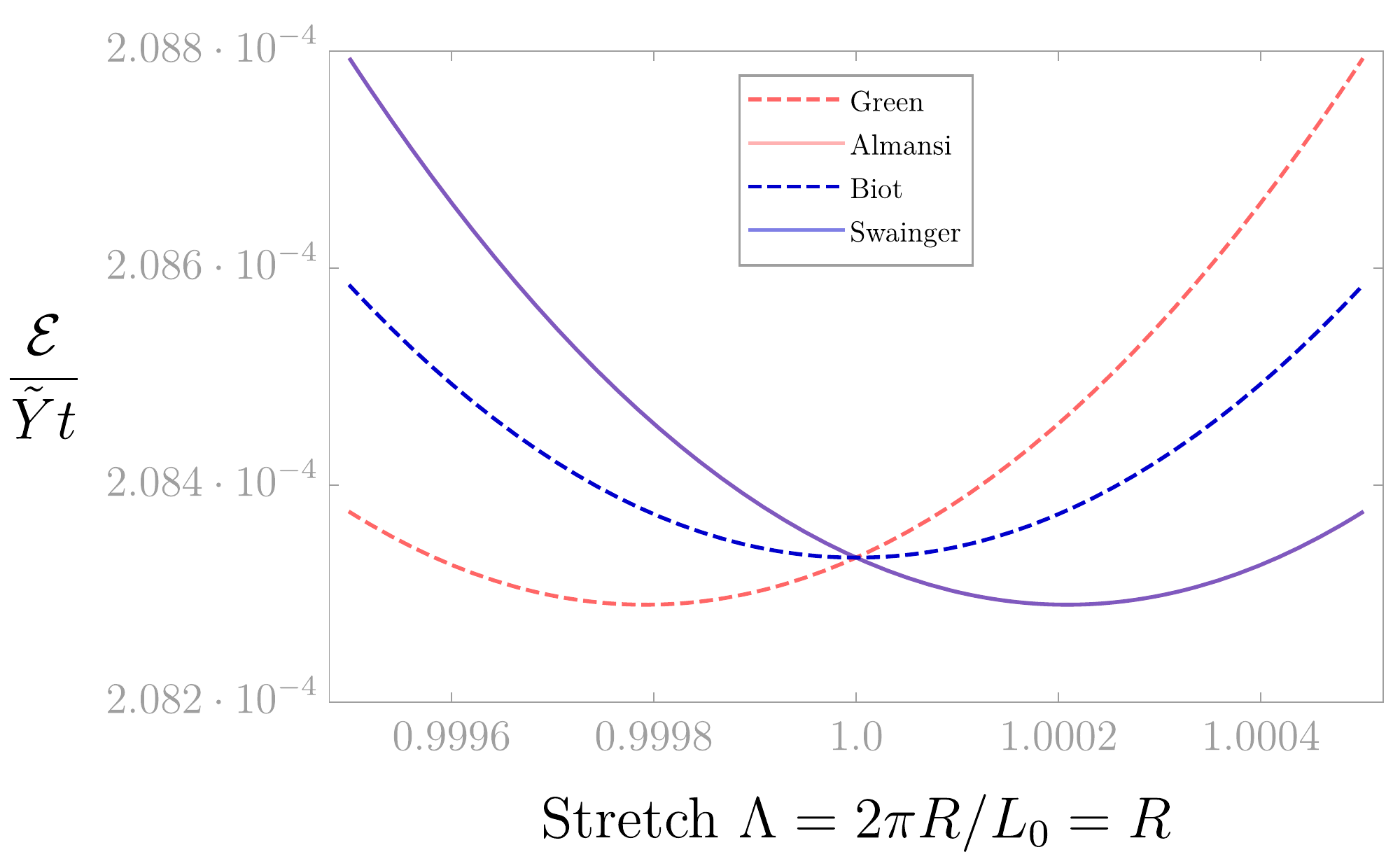}
	\caption{Total energy densities $\mathcal{E}$ as a function of ring radius $R$ for the closed ring of Figure \ref{ExpandingRadiusFigure}a with thickness $t = 0.05$ and $L_0 = 2\pi$.
The difference between Almansi and Swainger energies is not visible in this range.
These and Green experience shifts from the equilibrium Biot stretch of order $t^2/12$.}
	\label{CombinedEnergyFigure}
\end{figure}

Returning to the closed ring of Figure \ref{ExpandingRadiusFigure}a, the competition between stretching and bending may be seen in the total energies of Figure \ref{CombinedEnergyFigure}, calculated for a mid-line isometric radius of unity and a thickness $t=0.05$.
Among the class of symmetric circular solutions, only the Biot energy predicts a mid-line isometry as the ground state.  Almansi and Swainger, which are close enough to be indistinguishable in the small stretch range of the figure, relieve bending energy through an expansion of the ring to a larger radius, while Green contracts the ring.  These shifts in the equilibrium stretch scale as the ratio of bending to stretching moduli, $t^2/12$, with corresponding decreases in the energy scaling as the square of this ratio.
Explicitly, the minima are located at $\Lambda = \left(1 - t^2/6\right)^{1/2}$ for Green, $\Lambda = \left(1 - t^2/6\right)^{-1/2}$ for Almansi, $\Lambda =1$ for Biot, and $\Lambda =  1 + t^2/12$ for Swainger.

\clearpage

\section{Concluding discussion}\label{discussion}

The Biot-Antman definition of pure stretching is that which remembers and respects the original symmetry of the plate.  Changing the radius of a section of a bent, naturally flat body, while preserving its subtended angle, changes the lengths on either side of the mid-line by the same amount.  A suitable modification of this definition should be possible for naturally curved bodies.
Extension of our brief calculation to naturally curved beams and shells will likely prove informative.

All four quadratic energies have the same definition of pure bending, a change at constant stretch $\Lambda$ in the kinematic variable $\partial_x\theta$, the material derivative of the angle.  But only the Biot bending energy is simply the square of this variable, which should not be confused with the invariant curvature $\partial_x\theta/\Lambda$.  The other energies could, however, be rewritten as Biot plus a small correction.


Our simple example of a closed ring of naturally flat plate material illustrates important differences between the limits of vanishing strain (mid-surface isometry) and vanishing thickness.  
The effects we have noted vanish with thickness, not with strain.
At finite thickness, there are nontrivial effects of the choice of bending energy near isometry.  In particular, only the Biot strain predicts an axisymmetric isometric ring as the ground state, despite the absence of any incompatibility between the mid-line metric and the boundary conditions. 
This is an important point, given that a primary interest in soft matter elasticity is the manner in which (self-)incompatibility of two-dimensional metrics causes stretching and bending of sheets.
Additionally, those who seek to rationalize the behavior of soft plates and shells of finite thickness by invoking the concept of isometry might find a Biot model more agreeable
than the commonly employed Green model.
Finally, the simplicity of expression \eqref{Biotstrain} and exactness of the derivation \eqref{Biotenergy} points to the possible role of stretch as a fundamental quantity, with its deviation from unity being the appropriate choice for expansions, rather than more complex fields such as the metrics that appear in Green and Almansi.

Thus we should
reemphasize the potential value of a fully general Biot theory
whose development was not required by the present simple example.
Within the soft matter literature, Oshri and Diamant have recently developed an axisymmetric Biot theory \cite{OshriDiamant17}, and Knoche and Kierfeld \cite{KnocheKierfeld11} have worked with Biot-like measures of bending in shells.
A Biot theory would also connect with common bead-spring models \cite{SeungNelson88}.
The main hindrance appears to be the present lack of a simple and physics-friendly representation of a Biot energy in terms of position derivatives in arbitrary coordinates.  
An incompatible model would also require the concept of a reference stretch tensor akin to the reference metric employed in \cite{Efrati09jmps}.

\section*{Acknowledgments}

JAH would like to acknowledge a particularly helpful early conversation with S. S. Antman on this and closely related topics.  We also acknowledge helpful input on an earlier draft from P. Plucinsky and C. D. Santangelo.

\bibliographystyle{unsrt}


\end{document}